\documentclass[10pt,final]{article}

\usepackage{amsmath}

\makeatletter

\def\endequation{\eqno \hbox{\@eqnnum}$$\global\@ignoretrue}
\makeatother

\usepackage{amsfonts}
\usepackage{amsthm}
\usepackage{latexsym}
\usepackage{subfigure}

\usepackage{mydefines}

\usepackage{psfig}
\usepackage{epsfig}
\renewcommand{\mpar}[1]{}

\def\argmax{\operatornamewithlimits{arg\,max}}

\title{Effective Equations for Discrete Systems: \\ 
       A Time Stepper Based Approach }
\author{ 
Joakim M\"oller\thanks{Deparment of Numerical Analysis and Computer Science, KTH, 10044 Stockholm, Sweden.} \and
         Olof Runborg$^*$ \and
         Panayotis G. Kevrekidis\thanks{
Department of Mathematics and Statistics, University of Massachusetts,
Amherst, MA~01003 USA.}
\and Kurt Lust\thanks{
Departement Computerwetenschappen,
Katholieke Universiteit Leuven,
Celestijnenlaan~200A,
B-3001~Heverlee,
Belgium.
}
 \and 
Ioannis G. Kevrekidis\thanks{
Department of Chemical Engineering, 
Program for Applied and Computational Mathematics, Department of Mathematics, 
Princeton University, Princeton, NJ 08544, USA.} \and
	   }
\date{\today}

\begin{document}

\maketitle

\begin{abstract}

We propose a computer-assisted approach to studying
the effective continuum behavior of 
spatially discrete evolution equations.
The advantage of the approach is that the ``coarse model" (the continuum,
effective equation) need not be explicitly constructed.
The method only uses a time-integration code for the discrete problem and
judicious choices of initial data and integration times; our
bifurcation computations are based on the so-called Recursive Projection 
Method (RPM) with arc-length continuation (Shroff and Keller, 1993).
The technique is used to monitor features of the genuinely discrete
problem such as the pinning of coherent structures and its results are
compared to quasi-continuum approaches such as the ones based on 
Pad{\'e} approximations.

\mbox{}

\end{abstract}


\section{Introduction}
\lbsec{Intro}

In contemporary science and engineering modeling many situations arise
in which the physical system consists of a lattice of discrete interacting
units. 
The role of  discreteness in 
modifying the behavior of  solutions of continuum nonlinear PDEs
has recently been increasingly appreciated. 
The relevant  physical contexts
can be quite diverse, ranging from the calcium burst waves in living
cells \cite{g1} to the propagation of action potentials through
the tissue of the cardiac cells \cite{g2} and from chains of
chemical reactions \cite{g3} to applications in superconductivity
and Josephson junctions \cite{g4}, nonlinear optics and waveguide
arrays \cite{g5}, complex electronic materials \cite{g6},
the dynamics of neuron chains or lattices \cite{g7,g8} or the
local denaturation of the DNA double strand \cite{g9}. 

Whether the phenomenon in question is the propagation of an excitation wave
along a neuron lattice, the electric field envelope in an optical waveguide
array, 
or the behavior of a tissue consisting of an array of individual cells,
we would often like to model the system through a ``coarse level" 
effective continuum
evolution equation that retains the essential features of the actual
(discrete) problem.
Typically computational modeling of such systems involves two steps: the derivation
of effective continuum equations, followed by their analysis through traditional
numerical tools. 
In this paper we attempt to circumvent the derivation of explicit (closed) 
continuum effective equations, and analyze the effective
behavior directly.
This is accomplished through short, appropriately initialized simulations of the
detailed discrete process, a procedure that we call the ``coarse time stepper".
These simulations provide estimates of the quantities (residuals, action of Jacobians, 
time derivatives, Fr{\'e}chet derivatives) that would be directly evaluated 
from the effective equation, had such an equation been available.
The estimated quantities are processed by a higher level numerical procedure 
(in this case, the Recursive Projection Method, RPM, of 
Shroff and Keller \cite{ShroffKeller:93}) which 
computes the effective, macroscopic behavior (in this case, traveling waves
and their coarse bifurcations). 
A more general discussion of the combination of 
coarse time stepping with continuum numerical techniques
beyond RPM can be found in \cite{GKT}.
We have recently demonstrated such an approach to the computation of the effective
behavior (in some sense, homogenization) of spatially heterogeneous
problems 
\cite{RunKev:01}. 
This paper constitutes an extension of this idea to spatially discrete problems.

The paper is organized as follows: we begin with a brief review of the coarse time stepper
for spatially discrete problems. 
We then discuss our illustrative problem (a front for a discrete reaction-diffusion
system) and its properties.
A description of our implementation of the coarse time stepper for
the bifurcation analysis of this particular problem is then presented,
followed by numerical results.
We conclude with a discussion of an alternative approach that involves
the 
derivation
of an explicit effective evolution equation (based on Pad{\'e} 
approximations), 
and of the scope and applicability of our method.

\section{A Coarse Time Stepper for Discrete Systems}
\lbsec{Coarse}
Consider a discrete system where each
unknown is associated with a point on
a lattice in space.
In the discussion here, we consider a one-dimensional regular
lattice for simplicity. 
Higher dimensional and/or possibly irregular, lattices
can be treated in a similar way.
We denote the unknowns $\{u_\ell\}$, with $\ell\in \Znumbers$,
and the corresponding points $\{x_\ell\}$, such that
$x_\ell = \ell\Delta x$, where $\Delta x$ is
the lattice spacing.
We assume that the system is governed by 
the ordinary differential equations 
\be{GeneralEvolution}
  \frac{du_\ell}{dt} = F(t,u_{\ell-n},\ldots,u_{\ell+n}), 
\qquad \ell\in\Znumbers,
\ee
where $n>0$ is an integer representing the range of interaction
%
%
between lattice points.
We want to describe this discrete system dynamics through a
continuous function $v(t,x)$ that models
the ``coarse'' behavior of the unknowns on the
lattice:
$$
         u_\ell(t)\approx v(t,x_\ell),\qquad \forall t,\ell,
$$
in some appropriate sense.
We denote $v$ the {\it coarse continuous solution} of \eq{GeneralEvolution}
and we assume that $n$ is not large and that 
there exists an effective, spatially continuous evolution
equation for $v(x,t)$ of the form
\be{effeq}
   v_t = P(t,v,\partial_x v,\ldots,\partial^{M}_x v),
\ee
for some $P$ and integer $M$.
Such an effective equation for $v$ 
should ``average over" the detailed discrete structure of
the medium; if there are no macroscopic variations of the discrete medium,
this equation should therefore be translationally invariant;
for the moment, we will confine ourselves to this case.
In terms of \eq{GeneralEvolution}, we can express this as:
if $F$ does not depend on $\ell$,
and if $v$ and $\tilde v$ are two solutions to the effective
equation \eq{effeq} satisfying $v(0,x) = \tilde{v}(0,x+s)$ for all $x$, then
$v(t,x) = \tilde{v}(t,x+s)$ for all time $t>0$, all $x$, and all shifts $s$.

It is interesting to consider what the result of integrating such an effective
equation with a particular, continuum initial condition $v_0(x)$,
would physically mean.
There clearly exists an
uncertainty in how such a continuum initial condition would be imparted to 
(sampled by) the lattice.
One way would be to set $u_\ell(0) = v_0(x_\ell)$, for all $\ell$,
but we could equally well set $u_\ell(0) = v_0(x_\ell+s)$ for
any $s\in[0,\Delta x)$.
There exists, therefore, a one-parameter uncertainty parametrized by
a continuous shift $s$.
Simulations resulting from different lattice samplings of the
same continuum initial condition could be quite different.
This is best illustrated by thinking of a single-peaked function 
as the continuum initial condition: the peak may lie precisely 
at a lattice point, or could fall in-between lattice points.
It is reasonable to consider as a useful effective continuum equation
one which takes into account all possible shifts of the initial condition
within a cell; in analogy with our earlier work \cite{RunKev:01}, 
we would like to
analyze an effective equation that would describe the expected 
result---taken over all possible shifts---of sampling the initial condition
by the lattice.

We will use the coarse time stepper approach to simulate
an effective equation like \eq{effeq}. 
In this setting, we
approximate
$v(t,x)$ by the coarse time stepper solution
$\tilde{u}(t,x)$ at discrete times $nT$, where $T$ is the {\em time
horizon} of the coarse time stepper.
Using the terminology of this framework, we do
the following steps, starting from a continuous
initial condition $v_0(x)=\tilde{u}(0,x)$.
\begin{figure}
  \centerline{\mbox{\psfig{figure=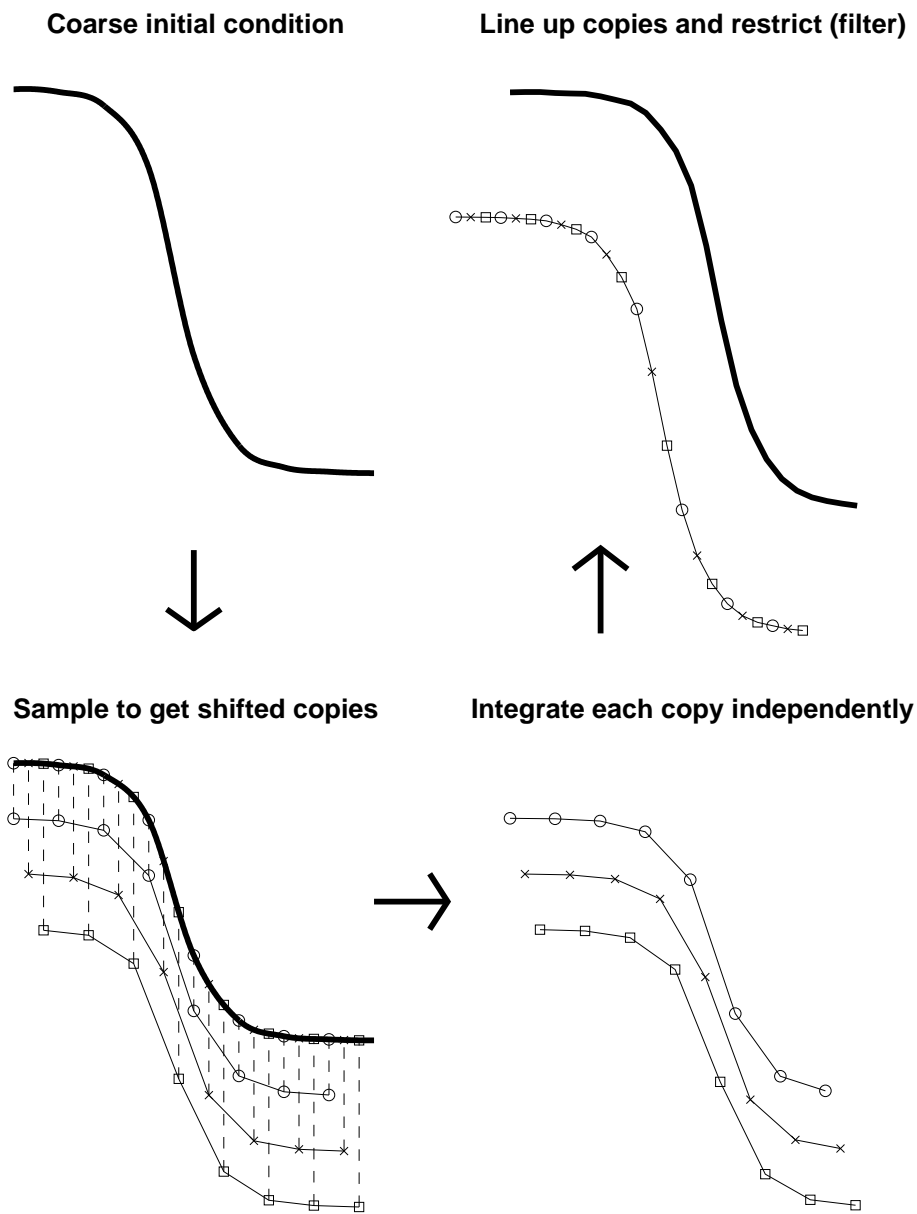,width=\textwidth}}}
  \caption{The coarse time stepper: Starting from a coarse initial condition $v_0(x)$, 
           lift it by sampling
           to an ensemble of initial data, $\{\u_j(0)\}$, $j=0,\ldots, N_c-1$, 
           for the system
            and evolve each set for time $T$.
           Line up solutions at time $T$ and interpolate to get
           $\bar{u}(x)$. Finally, filter $\bar{u}(x)$ to get $\tilde{u}(T,x)$,
           the result of the coarse time stepper at $t=T$.
}
  \lbfig{CTS}
\end{figure}

\begin{itemize}

\item {\em Lifting.} 
This initial data $v_0(x)$ is ``lifted" to 
an ensemble of $N_c$ 
different initial states of \eq{GeneralEvolution}
by sampling,
\be{LIFT}
u_\ell^j(0)=v_0(x_\ell+j\Delta s), 
\qquad \Delta s = \Delta x/N_c,
\qquad j=0,\ldots,N_c-1.
\ee
Setting $\u_j = \{u_\ell^j\}$, we
write this symbolically as
$$
   \u_j(0) = \mu_jv_0,
$$
where $\{\mu_j\}$ are called the lifting operators.
In this case they simply sample a continuous function.

\item {\em Evolve.} Each ensemble of initial data is
evolved till time $T$ according to the ``true dynamics" \eq{GeneralEvolution},
\be{EVAL}
{\u}_j(T)=\Tcal_{T}{\u}_j(0), \qquad j=0,\ldots,N_c-1.
\ee
where $\Tcal_\tau$ is the solution operator of \eq{GeneralEvolution} evolving ${\u}(t)$ to ${\u}(t+\tau)$.
This step thus generates an ensemble of solutions ${\u}_j(T)$ at time $T$. 

\item {\em Restrict.} 
Via the restriction operator $\Mcal$, the 
ensemble of solutions is brought back to a continuous function.
\be{REST}
\tilde{u}(T,x)={\Mcal}\{{\u}_j(T)\}, \qquad j=0,\ldots,N_c-1.
\ee
To ensure consistency we require that $\Mcal\{\mu_j\}=I$. 
The restriction operator $\Mcal$ is typically
defined as follows.
The solutions $\u_j(T)$ are thought of as
sample values of a function $\bar{u}$ such that 
$\bar{u}(x_\ell+j\Delta s)=u_\ell^j$. The
function $\bar{u}$ is recovered by interpolating
the sample values and
the restriction $\tilde{u}(x,T)={\Mcal}\{{\u}_j(T)\}$ is finally
given as a coarse scale filtering of $\bar{u}(x)$.
\end{itemize}
These steps are illustrated
in \fig{CTS}.
For  $n>0$ we define $\tilde{u}(nT,x)$ recursively
by applying the same construction. Hence,
\be{UtildeDef3}
\tilde{u}(nT,x)=\Mcal\{\Tcal_T\mu_j\}\tilde{u}((n-1)T,x).
\ee
The hope is that the coarse time stepper solution $\tilde{u}(nT,x)$,
at these discrete points in time, 
can be obtained from a closed evolution equation
like \eq{effeq} whose solution, $v(t,x)$ (defined for all $t$), 
agrees, at least approximately, with the coarse solution
obtained from the procedure above,
at the discrete points in time, $v(nT,x)\approx \tilde{u}(nT,x)$.
We will refer to the 
procedure as the {\em coarse time stepper}.

In order to approximate $v$ numerically, we
must use a finite representation of $\tilde{u}(nT,x)$. 
We let $\v^n=\{v_k^n\}_{k=0}^{M-1}$,
be this representation at time $t=nT$.
The elements $\{v_k^n\}$ could be nodal values,
cell averages or, more generally, coefficients for finite elements
or other basis functions. 
Let $\Pi$ be the operator realizing the
function from the finite representation, $(\Pi\v^n)(x)=\tilde{u}(nT,x)$.
We also require that the restriction operator
projects on the subspace spanned by the finite representation,
and we can redefine it to also convert the projected function
to this representation.
Symbolically, we then write the coarse time stepping
\be{COARSE}
\v^{n+1}=\Mcal\{\Tcal_T\mu_j\}\Pi\v^n =: G(\v^n).
\ee
Note that we may not be able to write 
down the explicit expression for $G$ or the equation \eq{effeq}
for $v(t,x)$, but our definition of $\tilde{u}(t,x)$
allows us to realize its time-$T$ map numerically in a straightforward
fashion.

Applied directly to the simulation, the coarse time-stepper does nothing
to reduce the cost of detailed computation with the discrete dynamics.
It is only in conjunction with other techniques (like projective 
integration \cite{gk}, or matrix-free fixed point techniques) that
the coarse time stepper may provide computational or analytical benefits.
Here we will make use of the coarse time stepper 
in conjunction with the Recursive Projection Method 
(RPM), to perform stability and bifurcation analysis of certain types
of solutions of the (unavailable) coarse evolution equation. 
For a schematic illustration of the coarse time stepper with RPM, see \fig{SCHEME}.
\begin{figure}[htb]
  \centerline{\mbox{\epsfig{figure=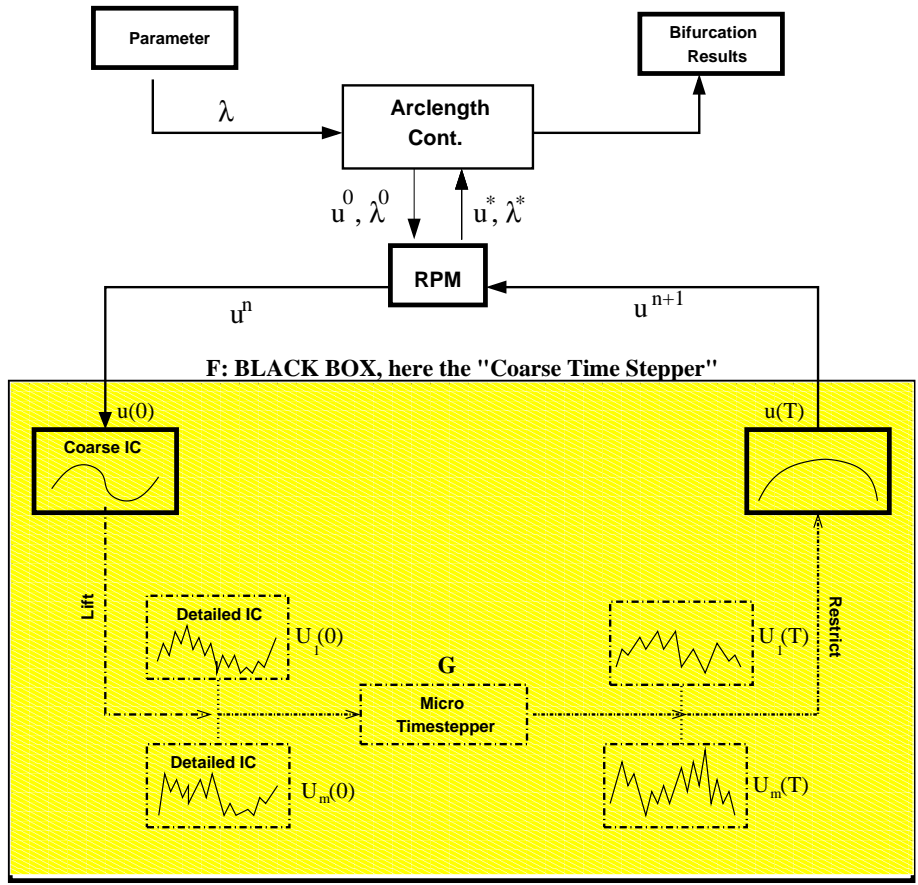,width=\textwidth}}}
  \caption{An overview of the coarse time stepper with RPM.}
  \lbfig{SCHEME}
\end{figure}

RPM helps locate fixed points, allows us to trace fixed point 
branches and locate their local bifurcations; 
when the bifurcations in \eq{COARSE} that we are
interested in do not involve fixed points, $G$ has to be reformulated.
How this is done depends on the application; 
for the type of solutions considered here (traveling fronts), 
the appropriate modification is discussed in \sect{Steady}.


\section{A Discrete Traveling Front Example}
\lbsec{DISCPROB}


The effects of discreteness on the propagation of traveling wave
solutions have been documented and analyzed in many different settings
over the last two decades. From the pinning of travelling waves
in discrete arrays of coupled torsion pendula and Hamiltonian models
\cite{IM,PK}, to the trapping of coherent structures in dissipative
lattices of coupled cells \cite{keener,fath} (see also references
therein), 
the role of spatial
discreteness has triggered a large interest in a diverse host of 
settings.

Effective equations capable of describing the nature of the solutions
of discrete problems should successfully capture the effects of discreteness
on the traveling wave shape and speed. 
More importantly, they should be capable of accurately predicting 
qualitative transitions (bifurcations) that are {\it 
inherently due to the discreteness}.
The most prominent of those is probably the pinning of
traveling waves and fronts often observed when the lattice spacing
becomes sufficiently large.
To illustrate the performance of our proposed coarse equation in
capturing such a front pinning, we have chosen what is arguably 
a prototypical spatially discrete problem capable of exhibiting it:
a one-dimensional lattice with scalar bistable on-site kinetics and
nearest neighbor diffusive coupling between lattice sites.
Our test problem is, therefore, a discrete reaction--diffusion system described by
\be{Model}
  \frac{du_{\ell}}{dt}=\frac{1}{(\Delta x)^2}(u_{\ell-1}-2u_\ell+u_{\ell+1}) + f(u_{\ell}), \qquad \ell \in \mathbb{Z}
\ee
with
\be{fDef}
 f(u) = 2u(u-1)(\eta-u), \qquad \eta=0.45.
\ee

This can serve as a model of e.g. individual cells in the cardiac tissue 
which are  resistively coupled through gap junctions (see e.g., 
\cite{keener} and references therein).
In this case the
solution $u_{\ell}$, would correspond to the electrical potential of the cells.
For small $\Delta x$ 
the system possesses solutions that can be characterized as {\it discrete traveling fronts}: 
see \fig{MOVE}. 
These solutions have a near constant shape and travel
in a ``lurching'' manner. 
When $\Delta x$  becomes sufficiently large, 
front propagation fails (front pinning). 
In our example, 
this happens at $\Delta x =\Delta x^*\approx 2.3$, see \fig{SPEEDintro}. 
The front speed for an infinite lattice approaches the asymptotic ``PDE speed"
value $0.1$ as the lattice size tends to zero.

\begin{figure}
  	\centerline{\mbox{\psfig{figure=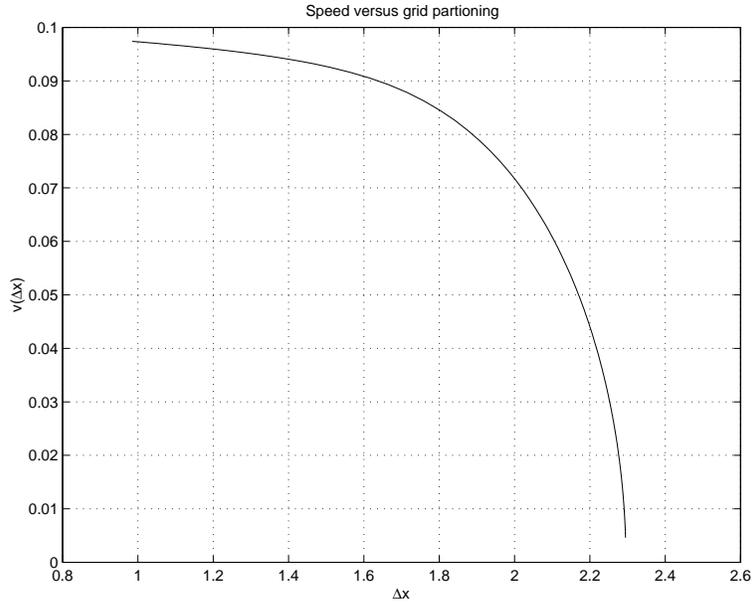,width=10cm}}}
	\caption{The speed of the front as a function of $\Delta x$. 
         As the lattice spacing is 
	increased, the speed $v$ approaches zero; the front stops
	at $\Delta x^* \approx 2.3$.  }
	  \lbfig{SPEEDintro}
\end{figure}
\begin{figure}
\centerline{
\mbox{
\subfigure[$u(t,x)$ in the $xt$-plane]{
  	\psfig{figure=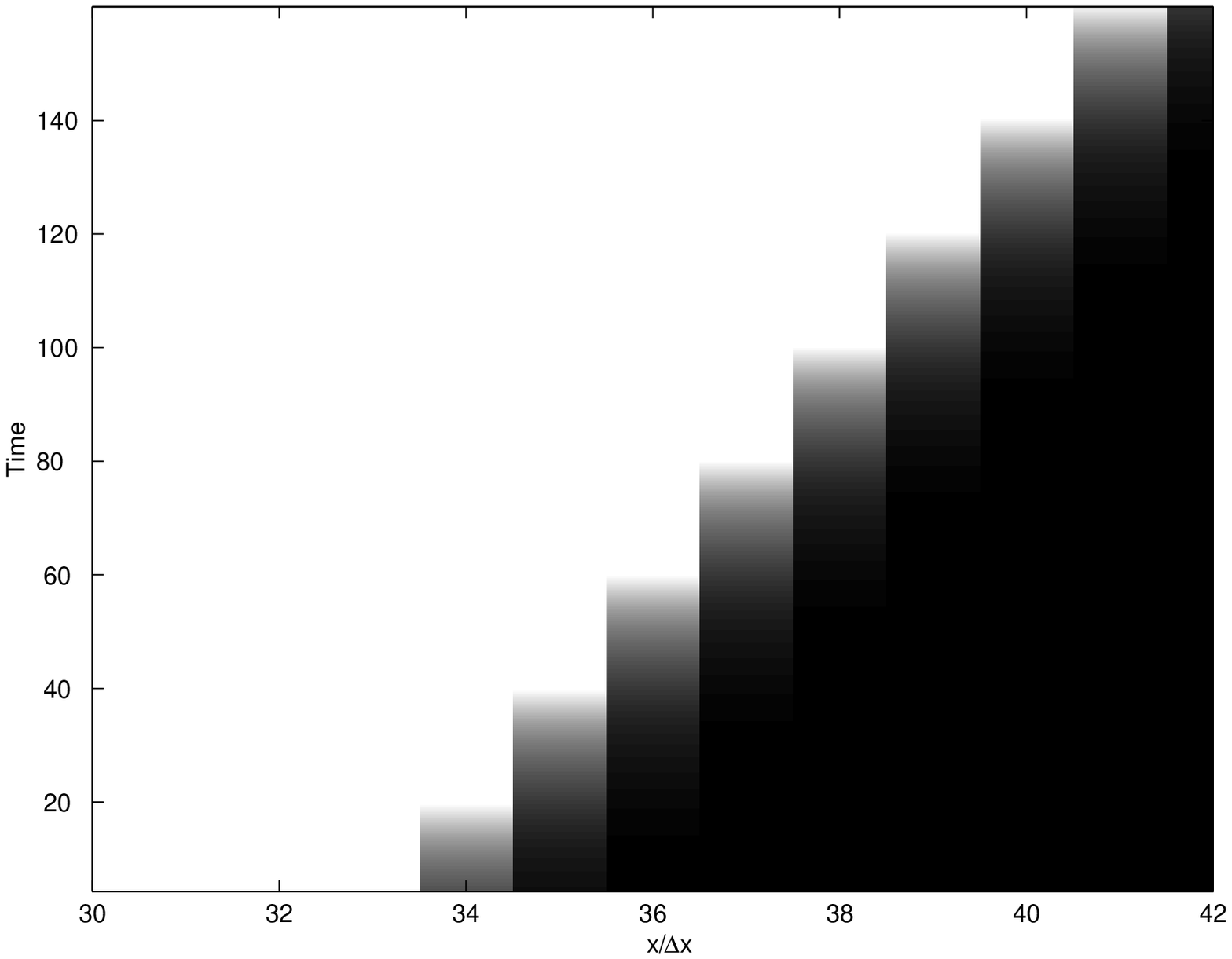,width=.5\textwidth}
}
\subfigure[$u(t,x)$, $t=0,\ 2.5,\ 5,\ \ldots,\ 40$]{
  	\psfig{figure=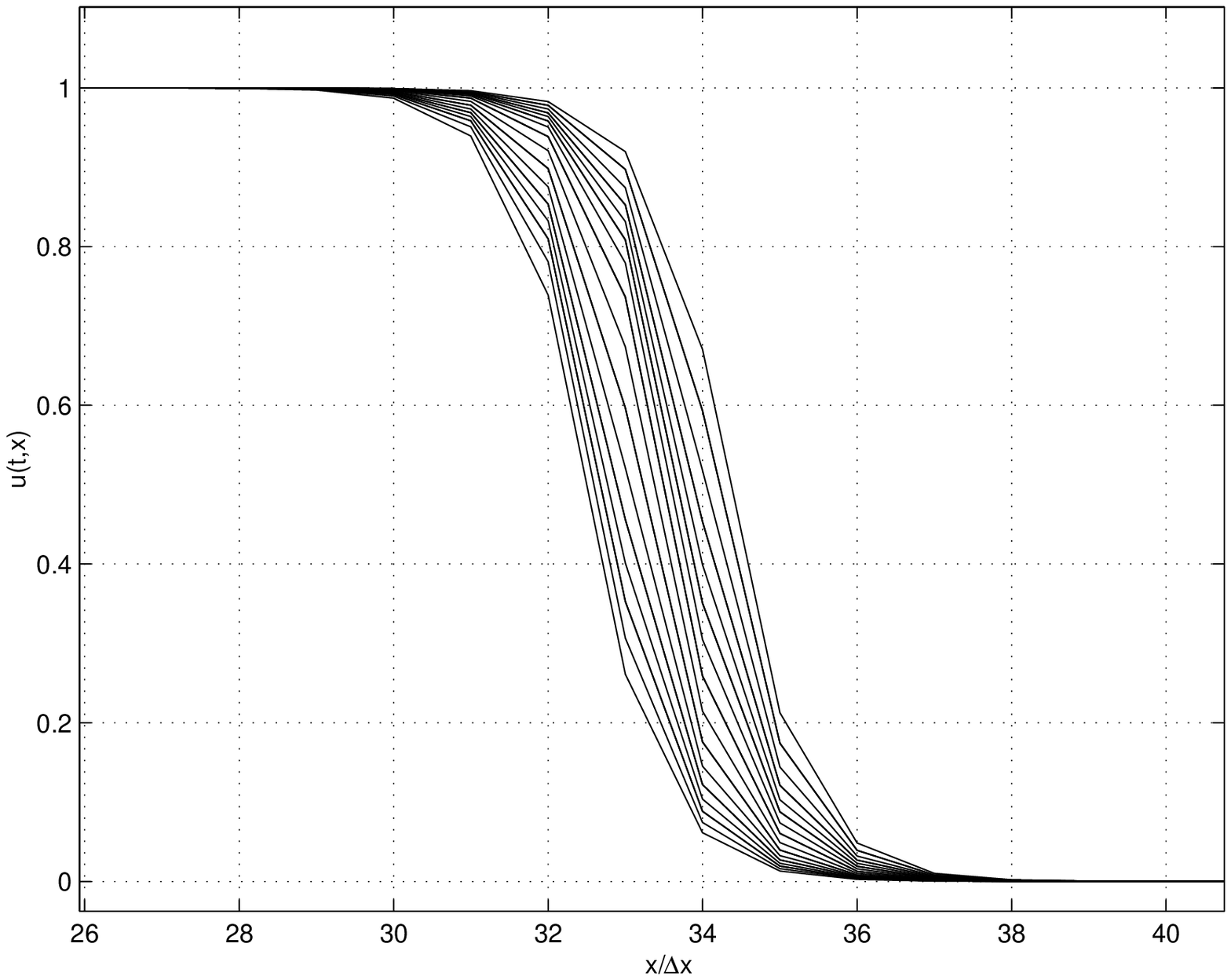,width=.5\textwidth}}
}
}
\caption{The plot illustrates how the front advances
when $\Delta x=1.75$.
The left figure shows the front in the  $xt$-plane; the
grayscale is proportional to the solution $u(t,x)$.
The right figure shows the solution as a function of $x$
at different time levels.
The time interval is $t\in[0,40]$. 
Looking at the spacing between the solution instances, 
we can see how the front speed varies in a lurching manner.}
	  \lbfig{MOVE}
\end{figure}

We will examine how faithful the coarse time stepper is
to the properties of
the solutions of the full discrete model \eq{Model}. 
Our numerical simulations are restricted to a finite domain, using $N=64$ grid points. 
At the boundaries, we prescribe Neumann-type conditions
\bea{BOUNDARY}
u_{N}-u_{N-1}&=&0, \nonumber\\
u_{0}-u_{-1}&=&0.   \nonumber
\eea
This should model the full problem accurately as long as the (relatively narrow) 
front is positioned sufficiently far from the boundary.

\subsection{Construction of the coarse time stepper}
\lbsec{Constr}

In this section we detail the procedures associated 
with the coarse time stepper applied to the test 
problem \eqtwo{Model}{fDef} on the finite interval $I=[0,L]$,
where $L=N\Delta x$ and the cell locations are $x_j=j\Delta x$, with 
$j=0,\ldots, N-1$.

Our choice of finite representation of the coarse
solution are $M$ nodal values $\v^n=\{v_k^n\}$, $k=0,\ldots, M-1$,
evaluated at $t=nT$ and $y_k=k\Delta y$, with $M\Delta y = N\Delta x$.

For many solution shapes
Fourier interpolation would be a natural interpolation operator
realizing the coarse solution $\tilde{u}(nT,x)$
from $\v^n$. We denote direct Fourier interpolation by $\Pi^f$.
We could then define the corresponding lifting operators
$\mu^f_j$ via the {\em shifting} operator $\Scal_s^f:\Real^M\to\Real^N$,
$$
   \mu^f_j\u := \Scal^f_{j\Delta s}\u,
\qquad
   (\Scal_s^f\u)_\ell := (\Pi^f\u)(x_\ell+s), \qquad
s\geq 0,
$$
where $\Pi^f$ uses $\{y_k\}$ as interpolation nodes.
In our case, however, the solution is not 
periodic on $I$ and we get large
errors if we use $\Scal_s^f$ directly. 
Instead we apply Fourier interpolation
to the {\em differences} of the $\v^n$ sequence.
We thus use the modified shifting operator
$\Scal_s:\Real^M\to\Real^N$ given by
\be{Sdef}
   \Scal_s\u :=C\Scal_s^fD,
\qquad (C\u)_\ell:=1+\sum_{j=0}^\ell u_j,
\qquad (D\u)_\ell:=\begin{cases}
 u_0-1, & \ell=0,\\
 u_\ell-u_{\ell-1}, & \ell>0.
\end{cases}
\ee
We then define
the lifting operator $\mu:\Real^M\to\Real^{N\times N_c}$
(acting directly on $\v^n$) as
$$
   \mu\v^n = \{\mu_j\v^n\},\qquad
\mu_j\v^n := \Scal_{j\Delta s}\v^n,
\qquad \Delta s = \frac{\Delta y}{N_c},
$$
where $j=0,\ldots,N_c-1$.

The restriction operator $\Mcal:\Real^{N\times N_c}\to \Real^M$
is also defined using the shifting operators, but
now with negative shifts,
$$
  \Scal_{-s}^f:\Real^N\to\Real^M,\qquad
   (\Scal_{-s}^f\u)_k := (\Pi^f\u)(y_k-s), \qquad
s\geq 0,
$$
where $\Pi^f$ uses $\{x_\ell\}$ as interpolation nodes.
We then set $\Scal_{-s}=C\Scal_{-s}^fD$ and let
$$
   \Mcal \{\u_j\} := \frac{1}{N_c}\sum_{j=0}^{N_c-1}\Scal_{-j\Delta s}\u_j.
$$
Note that these choices of $\mu$ and $\Mcal$
are consistent when $N\geq M$. 
Then, by the sampling theorem $\Scal^f_{-s}\Scal^f_s=I$
on $\Real^M$. Moreover, it is easy
to see that $CD=DC=I$. Therefore, we also have
$$
   \Scal_{-s}\Scal_s=
  C\Scal_{-s}^fDC\Scal_{s}^fD =
  C\Scal_{-s}^f\Scal_{s}^fD =
  CD = I,
$$
on $\Real^M$ and consequently,
$$
   \Mcal\mu \v^n = 
\frac{1}{N_c}\sum_{j=0}^{N_c-1}\Scal_{-j\Delta s}\mu_j\v^n =
\frac{1}{N_c}\sum_{j=0}^{N_c-1}\Scal_{-j\Delta s}\Scal_{j\Delta s}\v^n =
\frac{1}{N_c}\sum_{j=0}^{N_c-1}\v^n = \v^n.
$$
We should also remark here that, in the special case when $N=M$, we have
$$
\left(\frac{1}{N_c}\sum_{j=0}^{N_c-1}\Scal^f_{-j\Delta s}\u_j\right)_\ell
= (P_N\Pi^f\bar{\u})(x_\ell),
\qquad \bar\u=\{\bar{u}_r\}, \qquad \bar{u}_{\ell+jN}=u_\ell^j,
$$
where $P_N$ is a projection on the $N$ lowest Fourier modes.
Hence, if we used direct Fourier interpolation and $M=N$,
then our definition of $\Mcal$ is equivalent
to lowpass filtering of $\bar\u$, the lined up
copies described in \fig{CTS}, top right.
When we replace $\Scal^f_s$ by $\Scal_s$
we do not retain exactly this property, and
a definition of $\Mcal$ based on simple lowpass
filtering is no longer consistent. 
However, our procedure
still corresponds to a type of lowpass filtering,
although a more complicated one.

For the time integration of \eq{Model} we use
 the Crank--Nicolson method, treating the
 nonlinear term explicitly.
Thus, with $\w^0=\{w_\ell^0\}\in\Real^N$,
$$
  \Tcal_T\w^0 := \w^{N_T}=\{w_\ell^{N_T}\}, \qquad N_T\Delta t = T,
$$
where $\{w_\ell^n\}$ are given iteratively by
\bea{CN}
\lefteqn{
w_{\ell}^{n+1}-\frac{\Delta t}{2(\Delta x)^2}(w_{\ell-1}^{n+1}-2w_{\ell}^{n+1}+w_{\ell+1}^{n+1})}\hskip 15mm
\nonumber\\
&=&
w_{\ell}^{n}+\frac{\Delta t}{2(\Delta x)^2}(w_{\ell-1}^{n}-2w_{\ell}^{n}+w_{\ell+1}^{n})
+\Delta t f(w_{\ell}^n),\nonumber
\eea
for $\ell=0,\ldots,N-1$, together with the free boundary conditions
\bea{CNdummy}
w_{-1}^n-w_{0}^n&=&0,   \nonumber\\
w_{N}^n-w_{N-1}^n&=&0. \nonumber
\eea
In our computations we use the time step $\Delta t=0.01$.

\subsection{Steady state formulation}
\lbsec{Steady}

The coarse solution 
$\tilde{u}(nT,x)$ 
as we have defined it
is a (practically) constant shape moving front. 
In order to convert this moving state into
a stationary state, we can factor out the movement
through a procedure based on 
{\em template fitting} (\cite{RowleyMarsden:00,RunKev:01}, 
see also \cite{ChenGold:95})
which pins the traveling front at a fixed $x$-coordinate.
This is performed by a ``pinning-shift'' operator, which
we denote $\Pcal$. Our coarse time stepping is
then modified from \eq{COARSE} to
\be{COARSEMOD}
   \v^{n+1}=\Pcal\Mcal\{\Tcal_T\mu_j\}\Pi\v^n =:G(\v^n).
\ee
This formulation has a steady state at the
constant shape moving front.

Let us start from
the basic, Fourier based, pinning-shift operator 
$\Pcal^f:\Real^M\to\Real^M$. 
After introducing a template function $S(x)$, we define
\be{UbarDef}
   \Pcal^f\w := \Scal^f_c\w,\qquad
   c = \argmax_{c'\in\Real} \int_0^L ( \Pi^f {\w} )(x+c') S(x)dx.
\ee
Hence, $\Pcal^f\w$ is
the shifted version of $\w$ 
that best fits the template $S(x)$, in the sense that it
maximizes
the $L_2$-inner product between its Fourier interpolant and $S$.
Upon convergence, the effective front speed $v$ can be
deduced from the converged value of $c$ and the time reporting
horizon $T$ simply by taking $v=c/T$.
With the template $S(x)=1-\cos(2\pi x/L)$ we can compute
the inner product in \eq{UbarDef} explictly,
\be{OurShift2}
  \frac1L\int_{0}^L ( \Pi^f {\w} )(x+c') S(x)dx
 = \hat{w}_0-\Re \left(\hat{w}_1 e^{ic'}\right),
\ee
where $\hat{w}_k$ are the Fourier coefficients
of $\w$.
Hence, since $\hat{w}_0$ is real $c$ in \eq{UbarDef} should be chosen such that
$\hat{w}_1 e^{ic}$ is real and negative. This is easily implemented
numerically together with the Fourier shift $\Scal^f_c$.

For the same reasons as in the implementation of the
coarse time stepper, we would like to avoid direct
Fourier interpolation of the solution, since it
is not periodic. Therefore, we modify $\Pcal^f$
to operate on differences instead. In the same
spirit as in \sect{Constr}, we let
$$
   \Pcal :=C\Pcal^fD,
$$
with $C$ and $D$ defined in \eq{Sdef}. We still use 
the effective propagation speed given by $\Pcal^f$.

An important property of the Fourier based pinning
shift operator is that it satisfies $(\Pcal^f)^2=\Pcal^f$,
which follows from the sampling theorem  \cite{RunKev:01}.
For other types of interpolation, such as
piecewise polynomial interpolation, the pinning
shift operator will not have this property
and a steady moving coarse shape may not 
translate into a fixed point for \eq{COARSEMOD}.
Our modification still has this property though, since
$$
   \Pcal^2 = C\Pcal^fDC\Pcal^fD = C(\Pcal^f)^2D = 
   C\Pcal^fD = \Pcal,
$$
where we used the fact that $DC=I$.

\subsection{The RPM  with pseudo-arclength continuation}

RPM is an iterative procedure which can accelerate the
location of fixed points of processes; under certain conditions
it can help locate steady states of dynamic processes
(in particular, discretized parabolic PDEs).
It can be an acceleration technique for the solution of
nonlinear equations, and a stabilizer 
of unstable numerical procedures
(as it was first presented, \cite{ShroffKeller:93}).
Consider the fixed point problem
\be{FixedPointProblem}
   F(u;\lambda) = u,
\ee
and let $J$ be the Jacobian of $F$.

\begin{itemize}
\item Like the Newton method,
RPM can converge rapidly to the fixed point solution $u^*$
provided the initial guess is good enough;
the convergence occurs even if $J(u^*)$ has
a few eigenvalues larger than one.
The computational cost and convergence rate depend on
the eigenvalues of $J$. Optimally
there should be a clear gap in the spectrum
between small and large (near the unit circle)
eigenvalues and a limited number of
large (in norm) eigenvalues for RPM to perform well.

\item  $J$ never needs to be
evaluated directly, only $F$.
We can therefore apply RPM to any ``black box''
code that defines a function $F$; it is a ``matrix-free" method.
\item As a by-product, RPM also
computes approximations of the largest eigenvalues of $J$.
This gives approximate stability information about the fixed point.
\end{itemize}
When RPM is used for the computer-assisted bifurcation analysis
of steady states of (usually dissipative evolution) PDEs,
the function $F$ represents a {\it time-stepper}: a subroutine
that takes initial data and reports the solution of the
PDE after some fixed time (the reporting horizon $T$).
A fixed point then satisfies \eq{FixedPointProblem}.
The conventional way of finding
the steady state using a time-stepper would be to call it
many times in succession---in effect, to integrate the PDE for a
long time, corresponding to solving \eq{FixedPointProblem}
by simple fixed point (Picard) iteration.

RPM can improve this approach in two important respects.
First, the convergence can be significantly accelerated.
The nature of many transport PDEs usually encountered in engineering 
modeling
(the action of viscosity, heat conduction, diffusion, and the resulting
spectra)
dictates that there exists a separation of time-scales, which translates
into an eigenvalue gap in the spectrum of $J$ at the steady state.
Second, RPM converges even if the steady state
is slightly unstable, i.e. when $J$ has a few eigenvalues outside the
unit circle.
It may thus be possible to compute (mildly)
unsteady branches of the bifurcation diagram using forward
integration (but in a non-conventional way, dictated by
the RPM protocol).
RPM still retains the simplicity of the fixed point iteration,
in the sense that no more information is needed than
just the time-integration code.
This code, which may be a legacy code, and can incorporate the best
physics and modeling
available for the process, is used by RPM as a black box.

RPM can be seen as a modified version of fixed point
iteration. It adaptively identifies the subspace corresponding
to large (in norm) eigenvalues of $J$, hence the directions
of slow or unstable time-evolution in phase space.
In these directions the
fixed point iteration is replaced by (approximate) Newton iteration.
More precisely, suppose $F:\Real^N\times\Real\to\Real^N$
in \eq{FixedPointProblem}. Let $\Pbb$ be the maximal invariant
subspace of $J$ corresponding to the $m$ largest eigenvalues
and let $\Qbb$ be its orthogonal complement in $\Real^N$.
The solution $u$ is decomposed as $u=p+q=Pu+Qu$, where
$P$ and $Q$, are the projection operators
in $\Real^N$ on $\Pbb$ and $\Qbb$. These are constructed
from an orthogonal basis $V_p$
\bea{proj}
P&=&V_pV_p^T, \nonumber\\
Q&=&I-V_pV_p^T.\nonumber
\eea
In a pseudo-arclength continuation context the solution $u=u(s)$ and
$\lambda=\lambda(s)$, where $s$
parameterizes the bifurcation curve. 
In addition to  \eq{FixedPointProblem}
we then use
an algebraic equation to be able to handle turning points,
\be{ARC}
S(u,\lambda,\Delta s)=\frac{\|u(s)-u(s-\Delta s)\|^2}{\Delta
s}+\frac{|\lambda(s)-\lambda(s-\Delta s)|^2}{\Delta s}-\Delta s=0,
\ee
where 
$u(s-\Delta s)$ and $\lambda(s-\Delta s)$ refers to the converged
solution
at the previous point on the continuation curve.

The solution is advanced using a predictor-corrector method. Via
extrapolation from previous points  $u_i=u(s_i)$, $\lambda_i=\lambda(s_i)$
and $\Delta s_i=s_{i+1}-s_{i}$,
the predictor-solution is obtained. Comparing a first order
extrapolation,
\bea{predictor1}
\lambda^{*}&=&\lambda_i+\frac{\lambda_i-\lambda_{i-1}}{\Delta
s_{i-1}}\Delta s_i, \nonumber\\
u^{*}&=&u_i+\frac{u_i-u_{i-1}}{\Delta s_{i-1}}\Delta s_i, \nonumber
\eea
with a second order extrapolation,
\bea{predictor2}
\lambda^{**}&=&\lambda^{*}+
\frac{1}{2}\frac{\lambda_i(1-\gamma)-2\lambda_{i-1}+(1+\gamma)\lambda_{i-2}}{\Delta
s_{i-1}\Delta s_{i-2}}\Delta s_{i}^2, \nonumber\\
u^{**}&=&u^{*}+
\frac{1}{2}\frac{u_i(1-\gamma)-2u_{i-1}+(1+\gamma)u_{i-2}}{\Delta
s_{i-1}\Delta s_{i-2}}\Delta s_{i}^2 \nonumber\\
\gamma&=&\frac{\Delta s_{i-1}-\Delta s_{i-2}}{\Delta s_{i-1}+\Delta
s_{i-2}}, \nonumber
\eea
and requiring that
\be{CRIT}
\mathrm{max} (\|u^{**}-u^{*}\|,| \lambda^{**}-\lambda^{*} | )<\epsilon
\ee
the stepsize is determined. Here $\epsilon$ is a user specified
tolerance.
As the corrector method, we use RPM with pseudo-arclength continuation,
see \cite{ShroffKeller:93, Lust:97}. Starting from $u^0=u^{**}$
and $\lambda^0=\lambda^{**}$, the iterative scheme is given by
\bea{RPM}
q^{n+1}&=&QF(u^n,\lambda^n), \nonumber\\ \nonumber\\
\left[\begin{array}{ccc}
(V_p^TJV_p-I) & V_p^TF_{\lambda} \\
S^T_uV_p        & S_{\lambda}  \\
\end{array} \right]
\left[\begin{array}{cc}
\Delta p \\
\Delta \lambda \\
\end{array} \right]&=&-\left[\begin{array}{cc}
V_p^T F(p^n+q^{n+1},\lambda^n)-p^n \\
S(p^n+q^{n+1},\lambda^n) \\
\end{array} \right], \nonumber\\ \nonumber\\
u^{n+1}&=&p^n+V_p\Delta p^n +q^{n+1}, \nonumber\\
\lambda^{n+1}&=&\lambda^n+\Delta \lambda^n, \nonumber
\eea
where the left hand side consists of partial derivatives of 
$S$ in \eq{ARC} and of $F$ in
\eq{FixedPointProblem} with respect to $u$ and $\lambda$.
The iterates $u^n=p^n+q^n$ will converge to the solution
of \eq{FixedPointProblem} under the assumptions discussed above.
If the number of large norm eigenvalues, $m$,  is limited, the dimension
of $\Pbb$ and the projected Jacobian in the Newton iteration,
$V_p^TJV_p-I$, remains small.
Only this small matrix  needs to be
inverted. For a more complete description of RPM we refer to
\cite{ShroffKeller:93}.

\section{Numerical Results}

In this section we present some numerical results using
the coarse time stepper and the procedure described above
to simulate an effective equation for the discrete problem
in \eq{Model}.
We will start by discussing the ``exact" bifurcation diagram
of the discrete system, which we attempt to approximate.
We will then show results obtained through the coarse time stepper, 
and discuss the effect of time stepper ``construction parameters"
like the reporting time horizon, $T$ (the time
to which \eq{GeneralEvolution} is integrated within the coarse
time stepper), and the number of different initial shifted copies, $N_c$.

\fig{detbifdia} shows the bifurcation diagram of the discrete problem
as a function of the parameter $\Delta x$, the lattice spacing, in the
regime close to the onset of pinning. 
For lattice spacings smaller than $\Delta x^* \approx 2.3$ the system has,
as we discussed, an attracting,
front-like solution that travels; its motion
is {\it modulated} as it ``passes over" the lattice points.
For an infinite lattice, this modulated traveling solution possesses a
discrete translational invariance: $u_{\ell+1}(t+\tau) = u_\ell(t)$.
The shape of the modulating front is shifted by one (resp. $2, 3, \ldots, n$)
lattice spacing after time $\tau$ (resp. $2\tau, 3\tau, \ldots, n\tau)$; this helps
us define its effective speed $v(\Delta x) \equiv \frac{\Delta x}{\tau}$
(see \fig{SPEEDintro}).
As $\Delta x$ approaches zero, for an infinite lattice, the discrete front approaches
the continuum front of the PDE, and its speed (the period of the modulation
divided by $\Delta x$ approaches the PDE front speed, $0.1$ (see see \fig{SPEEDintro}).

If we identify shapes shifted by one lattice constant, the attractor
appears as a limit cycle with period $\tau$.
As the lattice spacing approaches the critical value $\Delta x^*$ the speed
of propagation approaches zero (the period of the ``limit cycle" approaches
infinity); asymptotically, $v(\Delta x) \approx |\Delta x - \Delta x^*|^{0.5}$.
As discussed in \cite{pla,carpio} what occurs is a Saddle-Node Infinite Period
(SNIPER) bifurcation: a saddle-node bifurcation where both new fixed points appear
``on" the limit cycle.
For larger values of $\Delta x$ the ``saddle" and the ``node" move away from each
other, and what used to be the limit cycle is now comprised from the saddle, the
node, and both sides of the one-dimensional unstable manifold of the saddle, which asymptotically
approach the node.

The saddle and the node are, of course, stationary fronts. 
A pair of them exists for every ``unit cell": all ``node fronts" are shifts of
each other by one lattice spacing, and all ``saddle fronts" are also shifts of each
other by one lattice spacing.
Since the medium has a discrete translational invariance, this makes 
sense---if an initial condition gives rise to a front eventually pinned at some location
in the discrete medium, the shift of this initial condition by one lattice spacing
will eventually get trapped one lattice spacing further.
This saddle-node bifurcation can be seen in \fig{detbifdia}a; linearizing around the
saddle front will give a positive eigenvalue $\lambda_s$, while the corresponding
eigenvalue $\lambda_n$ for the node front would be negative.
Since we look at the problem in discrete time, what is plotted 
is the {\it multiplier} $\mu_{n,s} = \exp(\lambda_{n,s}T)$, 
where $T$ is the reporting horizon.
The saddle front has a multiplier larger than 1, while the corresponding
multiplier for the stable node is less than 1; both multipliers asymptote
to 1 at the SNIPER ($\Delta x^*$).

\fig{detbifdia}b shows the bifurcation diagram in terms of the front traveling speed.
Since both the saddle and the node fronts are pinned (have zero speed)
they both fall on the zero axis; we plotted their eigenvalues in \fig{detbifdia}a to
distinguish between them.
The true traveling speed (broken line) is compared with the effective traveling speed
predicted by a coarse time-stepper using $N_c=5$ copies within each unit cell,
and a reporting horizon of $T=32$.
The coarse time stepper speed is a byproduct of fixed point computation and
continuation with it; short bursts of detailed simulation are used in the RPM
framework to construct a contraction mapping that converges to a fixed point of the time stepper.
The final shift upon convergence (from the pinning-shift computation), divided by
the time stepper reporting horizon gives us an estimate of the ``effective speed".
Inspection of \fig{detbifdia}b indicates that the coarse time stepper never predicts a
speed that is exactly zero; yet it gives a good 
approximation of the effective speed, all the way from small $\Delta x$ to the near
neighborhood of the pinning transition, when the effective speed becomes small.

We will return to discussing this issue of ``small residual motion" 
for the coarse time stepper shortly.
To give an indication of when the
procedure stops being quantitative, we have included the $\Delta x/T$ curve in \fig{detbifdia}b: 
disagreement starts well in the regime 
where the effective movement is {\it less} than one unit cell per observation
period.
In the next section we will compare the ``goodness of approximation" of 
our coarse time stepper to the effective speed predicted by the Pad{\'e} approach
to extracting effective continuum equations.
It is interesting that the coarse time stepper sometimes predicts a small hysteresis
loop at low speeds, relatively close to ``true pinning"; notice in \fig{detbifdia}a the 
unstable (larger than one) multipliers for the brief saddle part of this loop.
We will discuss a tentative rationalization of this below.

\fig{diffParam} illustrates the effects of ``time stepper construction" parameters
on the effective behavior predicted by the time stepper: the reporting time-horizon,
for two different sets of shifted copies ($N_c=3$ and $N_c=10$) as well as the effect of the
number of copies for a fixed time horizon ($T=16$).
Augmenting the time stepper reporting horizon is shown in 
\fig{diffParam}a-b;
clearly, in both cases, extending the time stepper reporting horizon
extends the region over which its effective speed agrees with the true problem
closer to $\Delta x^*$. Larger numbers of copies ($N_c=5, 10, 20$) also perform
slightly better than smaller numbers ($N_c=3$).
In all cases the qualitative behavior is the same: (a) successful
approximation of the effective speed until reasonably close to 
true pinning; (b) all differences occur when the average front motion is
significantly less than one unit cell per reporting horizon; (c) there is always
a slight residual motion, which---possibly after a small hysteresis loop close to 
true pinning---eventually becomes negligible.

\begin{figure}[htbp]
\centering
\subfigure[Multipliers versus $\Delta x$]
{\psfig{file=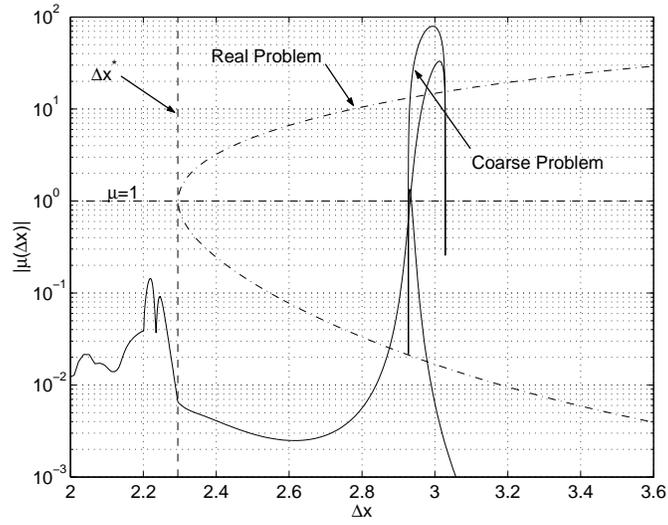, height=7cm}}
\subfigure[Effective front speed versus $\Delta x$]
{\psfig{file=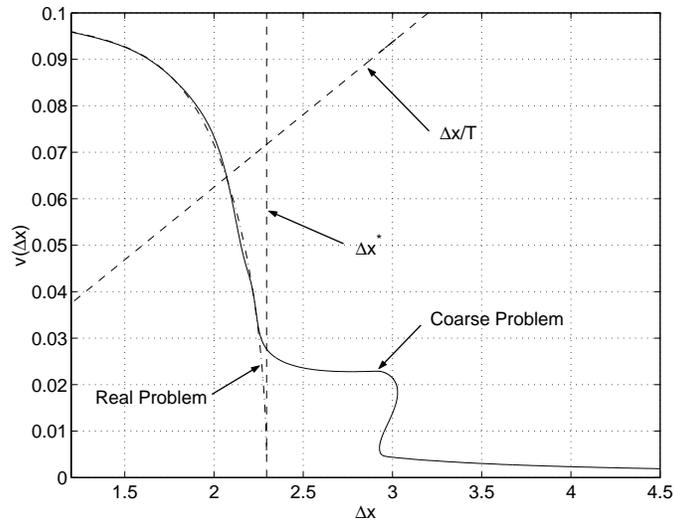 , height=7cm}}
\caption{Detailed bifurcation diagram and coarse time stepper
bifurcation diagram with parameters $N_c=5$, $T=32$.}
\lbfig{detbifdia}
\end{figure}

\begin{figure}[htbp]
\centering
\mbox{
\subfigure[Varying time horizon, $T=10,\ldots,50$, with fixed $N_c=3$.
           Dashed lines show $\Delta x/T$,
           i.e. speed required to traverse one cell.]
{\psfig{file=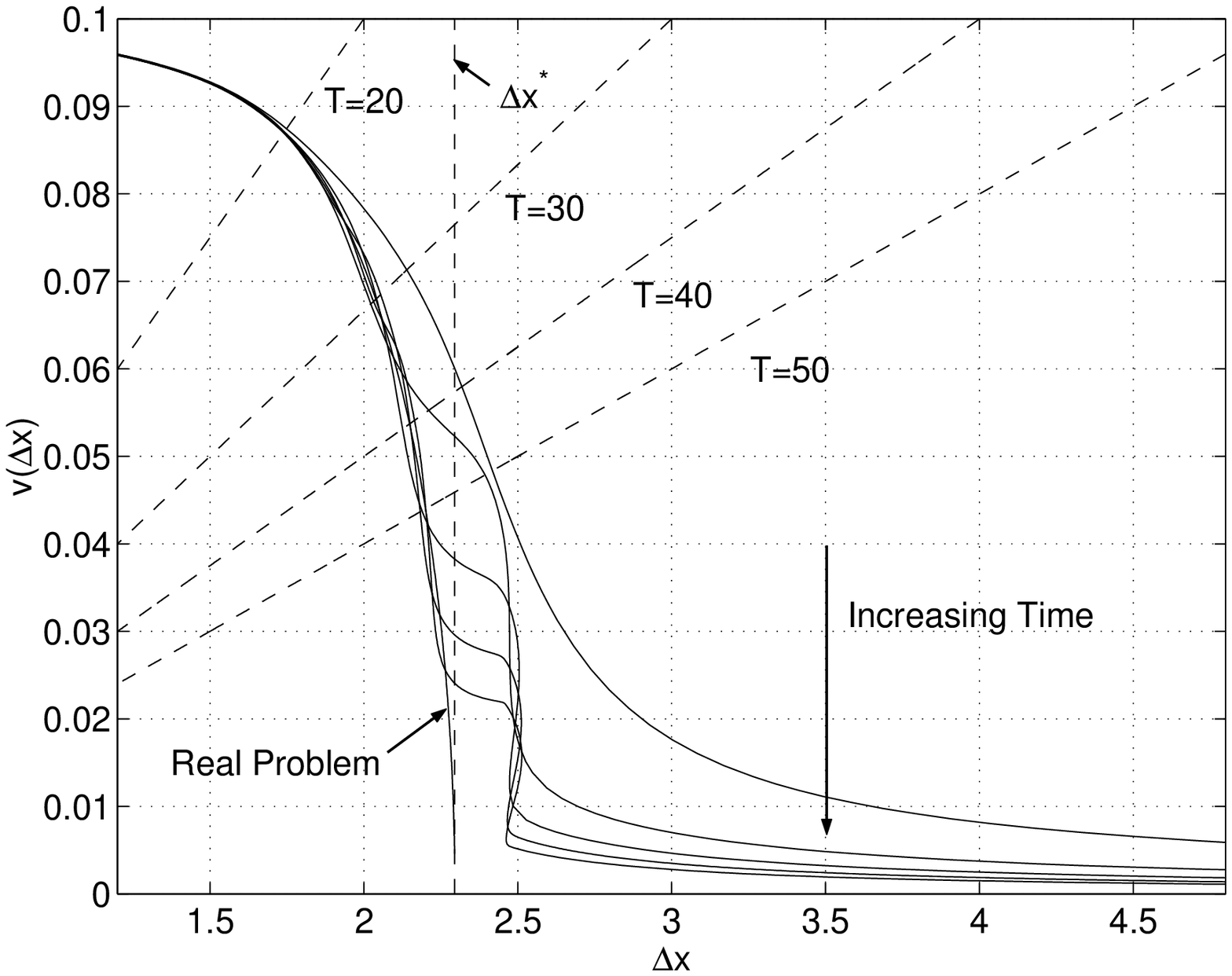, width=0.5\textwidth}}
\hskip 5mm
\subfigure[Varying time horizon, $T=5,\ldots,16$, with fixed $N_c=10$.]
{\psfig{file=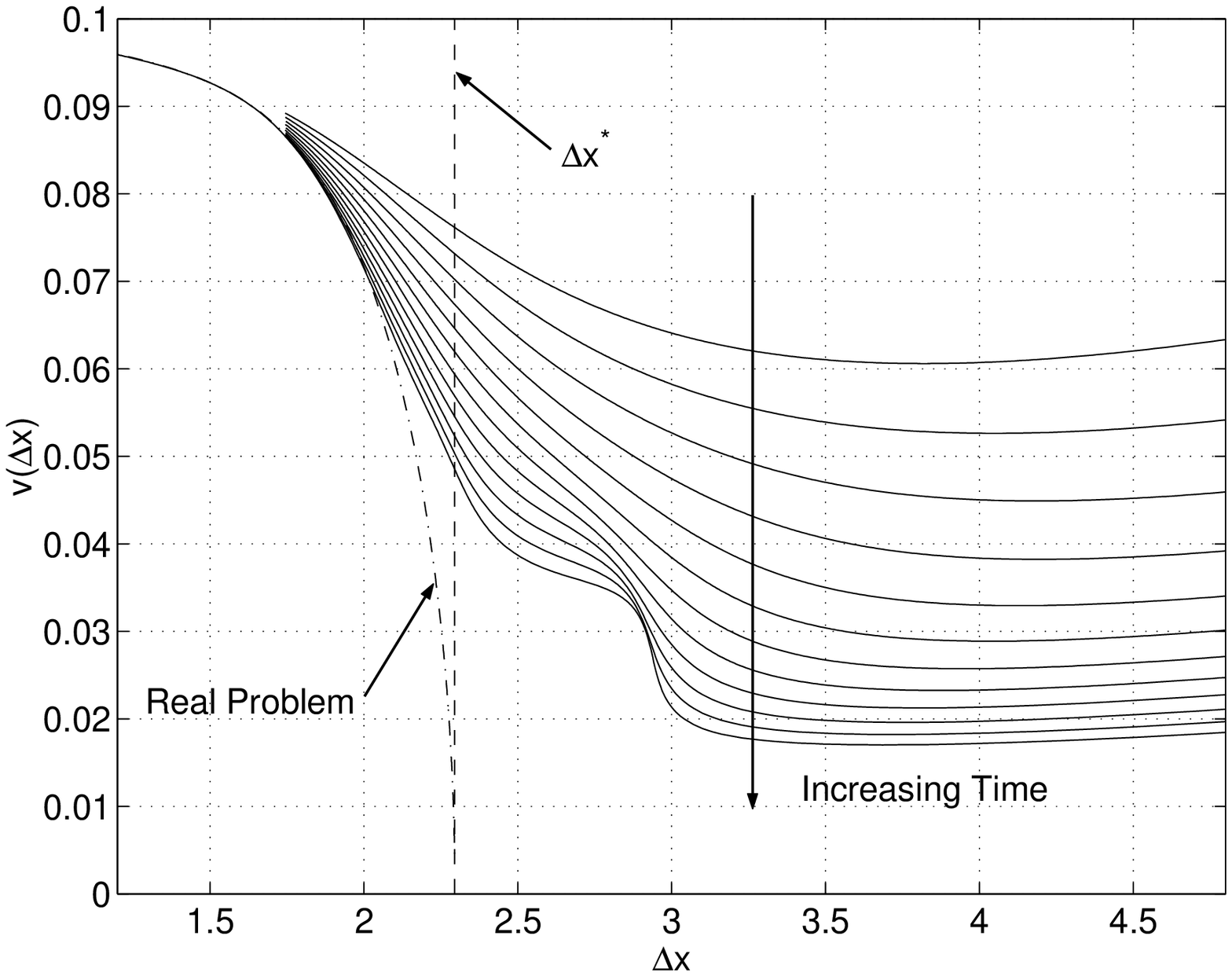, width=0.5\textwidth}}
}
\subfigure[Varying number of copies, $N_c=3,5,10,20$, with fixed $T=16$.]
{\psfig{file=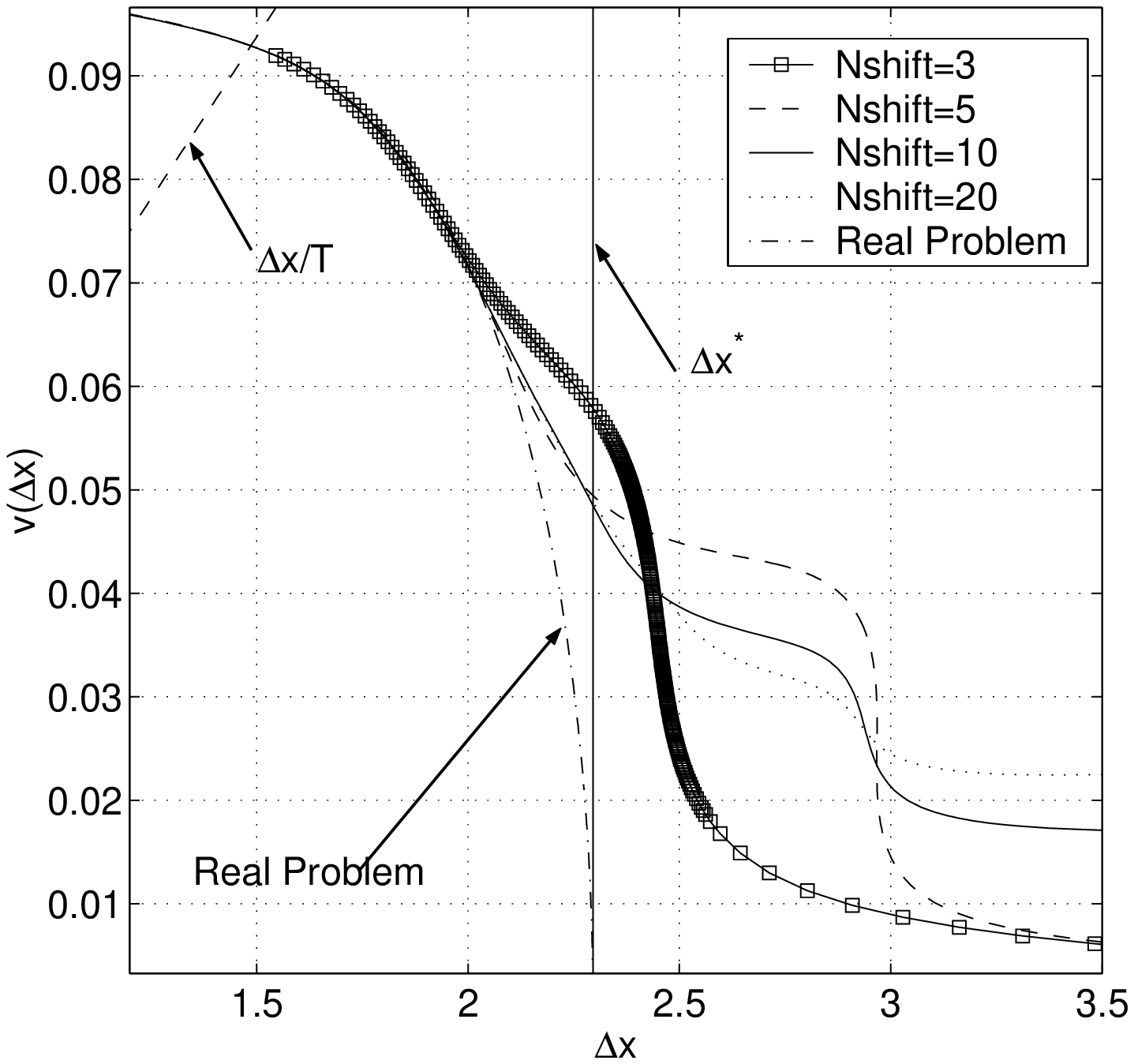, width=0.5\textwidth}}
\caption{Effective front speed versus $\Delta x$
         and the effect of varying the time horizon $T$
         and the number of copies $N_c$.
}
\lbfig{diffParam}
\end{figure}

We now turn to the discussion of the slight residual motion of the
coarse time stepper at large $\Delta x$ beyond $\Delta x^*$.
For an infinite domain, the saddle and node pinned fronts appearing there
are invariant to translations by one lattice
spacing; for a large enough computational domain we still see two pinned
front solutions per cell.
When we ``sprinkle" initial conditions along the cell, depending on their 
location with respect to the saddle front, the trajectories may either be attracted 
to the stable node ``to the right" or to the one ``to the left" of the saddle.
It is instructive to represent these solutions as in \fig{copymovement}a,
in a way that identifies the ``right"
node front with the ``left" one; here translation along the lattice 
corresponds roughly to rotation along the circle. 
The node is denoted by a black circle, and
the saddle by a white one. The small squares represent the
initial positions of our initial condition ``copies".
The fate of our distribution of initial conditions is governed by their initial
``angle" on the circle---as our time horizon grows all initial conditions will
asymptote to a stable front, either the left one 
(moving counterclockwise on the circle) or the right one (clockwise movement). 
We now see clearly the physical reason behind the net residual motion for any finite time
horizon for the coarse time stepper.
An initial condition that is put down ``at random" in a unit cell deep in
the pinned regime, even if it never exits this unit cell, will 
gradually traverse
the part of the circle separating it from the closest node front.

When the critical parameter value is approached from the pinned side, the saddle and the 
node fronts approach each other on the circle, on their way to 
coalescing at the SNIPER bifurcation point \cite{pla,carpio}.
\fig{copymovement}b shows how this process becomes manifest in 
the coarse time stepper computations, using the problem in \fig{detbifdia} as
our example.
Deep in the pinning regime (high $\Delta x$, marked $\alpha$) 
the relative ``phase" of the saddle
and the node pinned fronts on the circle remains roughly constant. 
The distance each member of our ensemble of initial conditions has
traversed during one time horizon can be
deduced from \fig{copymovement}b: the copy with the largest negative 
movement is
the one closest to the saddle but on its left (copy number two).
One can similarly rationalize the labelling of the remaining curves in
\fig{copymovement}b.
When $\Delta x$ is reduced approaching the onset of pinning, at some point the
saddle front starts moving appreciably towards the node front. 
As part of this movement, it ``sweeps" the circle counterclockwise; at $\Delta x \approx 2.8$
it has its first encounter with one of our initial conditions---the closest one on the left.
When the saddle ``moves past" it into the regime marked $\beta$, 
this copy, which was responsible for the largest negative
displacement now approaches asymptotically the node front on the right, performing the largest
{\it positive} displacement (and so on for the remaining copies).
Eventually, in the propagating regime, marked $\gamma$, and for long enough reporting horizons, the initial ``phase"
difference (a fraction of a cell) becomes negligible compared to the net displacement of
each point (several cells).

The real movement in phase space is shown un
\fig{copymovement}c for two different $\Delta x$.
In these subfigures, the $x$-axis represents
$\sin(2\pi x_c)$ where $x_c$ corresponds to the location of
the front, more specifically $x_c=\sum_\ell(D\u)_\ell\ell$.
The $y$-axis represents $\max_\ell |(D\u)_\ell|$. The
initial positions of the copies are indicated by
small squares and their locations at $t=T$, the time horizon,
are marked by filled circles. 
The labels refer to the same copies as in
\fig{copymovement}b.

As the reporting time horizon of the time stepper goes to infinity, it is clear
that one can compute the average residual movement from the asymptotic 
position of the saddle front, i.e. from the relative extent of
the circle ``to the right" and ``to the left" of the saddle front.
The most reasonable point to ``declare" as an estimate of the true pinning
from coarse time-stepper computations would come from a polynomial extrapolation
of the ``successful" regime (close to the tip of the ``apparent parabola" in
\fig{detbifdia}); alternatively, a value of $\Delta x$ where the speed is small enough
(well below one unit cell per time horizon) and its variation with number of copies
and time horizon is below a user-prescribed tolerance, would also serve this purpose.
While there is no well defined pinning bifurcation for the coarse
time stepper (since pinning is an inherently non-translationally invariant bifurcation),
the procedure can provide a good approximation of the effective
shape and speed of the traveling fronts, as well as ``common sense" ways of 
numerically estimating the true pinning. 

\begin{figure}[htbp]
\centering
\subfigure[Schematic movement of copies in phase space]{\psfig{file=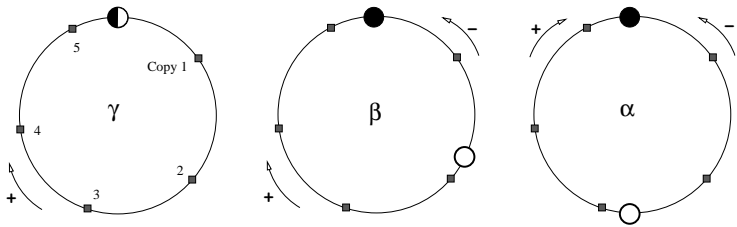 , width=0.8\textwidth}}
\vskip -3mm
\subfigure[Distance traversed by copies]{\psfig{file=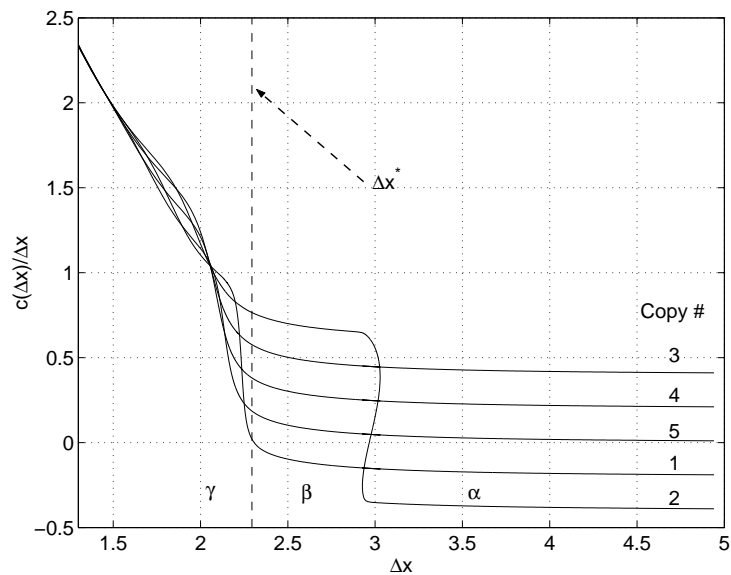,width=0.8\textwidth }}
\vskip -3mm
\subfigure[Real movement of copies in phase space for
$\Delta x=1.6$ (left) and $\Delta x=2.3$ (right).
(See text for specification of axes.)]{\psfig{file=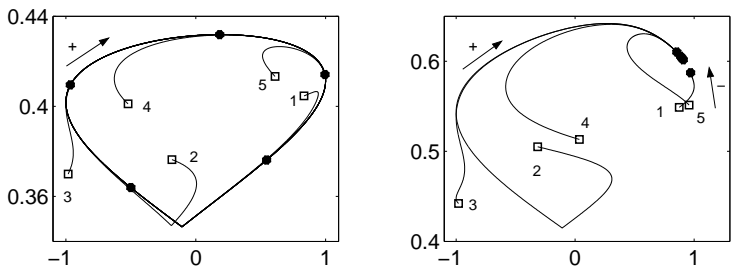,width=0.8\textwidth}}
\caption{Movement of the individual copies, for $N_c=5$, $T=32$. 
}
\lbfig{copymovement}
\end{figure}

\section{An Alternative Continuum Approach: Pad{\'e} Approximations}

In this section, we propose an alternative scheme for capturing
effects of discreteness, by means of a (now explicit) continuum
equation. This PDE is obtained by means of Pad{\'e} approximations \cite{p1,p2}
which can be used to approximate
discreteness in a quasi-continuum way, through the use of pseudo-differential
operators.
In particular, starting from the
Taylor expansion for analytic functions, see e.g., \cite{yug1},
\begin{eqnarray*}
u(x+m)=\exp(m \partial_x) u(x),
\label{geq2}
\end{eqnarray*}
one can then express spatial discreteness as 
\begin{eqnarray*}
u_{\ell+1}+u_{\ell-1}-2 u_\ell &\equiv& (\exp(\Delta x \partial_x)+\exp(-\Delta x \partial_x)-2)
u(x)
\nonumber\\
&\equiv & 4\sinh^2(\frac{\Delta x \partial_x}{2}) u(x,t).
\end{eqnarray*}
Expanding $\exp(\pm \Delta x \partial_x)$ \cite{p2}, one then obtains
\begin{eqnarray*}
\exp(\pm \Delta x \partial_x)-1=\frac{1}{2} \Delta x^2 (1+\frac{\Delta x^2}{12}
\partial_x^2 + \dots) \partial_x^2 \pm \Delta x (1+\frac{1}{6} \Delta x^2
\partial_x^2 + \dots) \partial_x
\label{geq4}
\end{eqnarray*}
Finally, regrouping the terms in the manner of Pad{\'e} \cite{p1,p2}
yields
\begin{eqnarray}
\exp(\pm \Delta x \partial_x)-1 \approx \frac{1}{2} 
\frac{\Delta x^2 \partial_x^2}{1-\frac{\Delta x^2}{12} \partial_x^2} \pm 
\frac{\Delta x \partial_x}{(1-\frac{\Delta x^2}{12} \partial_x^2)^2}
\label{geq5}
\end{eqnarray}
We now use the pseudo-differential operator 
approximation in (\ref{geq5}) to convert
the discrete model in \eq{Model}
into the PDE approximation of the form:
\begin{eqnarray}
u_t = \frac{ \partial_x^2}{1-\frac{\Delta x^2}{12}  \partial_x^2}
u + f(u).
\label{geq6}
\end{eqnarray}
Such approaches were introduced and used extensively 
by Rosenau and collaborators \cite{r1,r2,more} to regularize
nonlinear wave equations, particularly of the Klein--Gordon
type.

Eq. (\ref{geq6}) clearly emulates the discrete setting in some
key aspects of the relevant spectral operator properties
(i.e., of the discrete Laplacian in comparison with the pseudo-differential
operator of (\ref{geq6})). For example, considering plane wave
solutions of the form $ 
\exp(\lambda t-i k x)$, we obtain in the
discrete case the linearized dispersion relation (around a uniform state
$u=u_{\rm hom}$)
\begin{eqnarray*}
\lambda= \frac{2}{\Delta x^2}  \left(\cos(k \Delta x)-1) \right) + 
f'(u_{\rm hom}). 
\end{eqnarray*}
In the case of (\ref{geq6}), the corresponding equation becomes
\begin{eqnarray*}
\lambda=- \frac{k^2}{1+ \frac{\Delta x^2}{12} k^2} + f'(u_{\rm hom}).
\end{eqnarray*}
Apart from sharing the continuum limit, the two dispersion relations
share another qualitative feature which is particularly
important \cite{r1,r2,more}; 
namely, the presence of a lower bound in the continuous spectrum. 
Notice, however, that the two lower bounds are different 
($f'(u_{\rm hom})-4/\Delta x^2$ in the 
discrete case versus $f'(u_{\rm hom}) - 12/\Delta x^2 $ in the Pad{\'e} 
approximation).

It would then be of interest to 
alleviate this spectral discrepancy, as well as to match the discrete
operator (if possible) to a higher order in the Taylor expansion  
\begin{eqnarray*}
\frac{u_{n+1}+u_{n-1}-2 u_n}{\Delta x^2}=\sum_{j=1}^{\infty} 
\frac{2 \Delta x^{2 j-2}}{(2 j)!} u^{(2 j)}
=u_{xx}+ \frac{\Delta x^2}{12} u_{xxxx} + \frac{\Delta x^4}{360} u_{6x}
+ O(\Delta x^6).
\label{geq14}
\end{eqnarray*}

This can be achieved by a natural generalization in the form of a
continued fraction such as e.g., 
\begin{eqnarray}
\frac{\partial_x^2}{1-
\frac{A \partial_x^2}{1-\frac{ B \partial_x^2}{1- C \partial_x^2}}}.
\label{geq17}
\end{eqnarray}
In order to use (\ref{geq17}) in practice (i.e., for computational
purposes), we convert the three fractions into one of the form
\begin{eqnarray}
\frac{\partial_x^2(1 + \alpha \Delta x^2 \partial_x^2)}{1+ (\alpha+\beta) \Delta x^2 \partial_x^2 + \gamma
 \Delta x^4 \partial_x^4},
\label{geq19}
\end{eqnarray}
where a simple (algebraic) reduction of $A,B,C$ to $\alpha,\beta,\gamma$
has been used.
We then use Taylor expansion of the denominator to convert the
expression
of (\ref{geq19}) into one resembling (\ref{geq14}). By
matching
up to $O(h^6)$ the exact Taylor expansion, we obtain three algebraic
equations for $\alpha,\beta$ and $\gamma$. In this way, we obtain
a set of solutions for $\alpha,\beta$ and $\gamma$. We use here
the set $\alpha=-0.007$ $912$, $\beta=-1/12$, $\gamma=0.002$ $056$.
An additional benefit (to the matching of the Taylor
expansion up to correction terms of $O(h^8)$) that should be
highlighted here is the value $\alpha/(\gamma \Delta x^2)=3.848/\Delta x^2$ 
of the lower bound expression 
for $\lambda$, which is much closer to the theoretical
lower bound of $4/\Delta x^2$ than the prediction $12/\Delta x^2$ of the 
leading order approximation presented previously.
The resulting evolution equation will then read:
\begin{eqnarray}
u_t=\frac{\partial_x^2(1 + \alpha \Delta x^2 \partial_x^2)}{1+(\alpha+\beta) \Delta x^2 \partial_x^2 + \gamma
 \Delta x^4 \partial_x^4} + f(u)
\label{newmodel}
\end{eqnarray}

Both (\ref{geq6}) and (\ref{newmodel}) can be numerically 
implemented in a straightforward manner, by means of the spectral
techniques described in \cite{contdis}. 
We have performed numerical simulations of the front propagation,
using $1024$ modes in the spectral decomposition of (\ref{geq6})
and (\ref{newmodel}). We will refer to these equations as the 
(Pad{\'e}) models A and B respectively. 
A fourth order Runge--Kutta algorithm has been
used for the time integration.
For each value of $\Delta x$, we identify the position $x_c$ of the
front as the point where the ordinate of the front acquires the
value $u=1/2$. The linear interpolation scheme suggested in
\cite{boesch} has been implemented  and has proved to be an efficient 
front tracking algorithm in all the examined cases.

\begin{figure}[htbp]
\centering
{\psfig{file=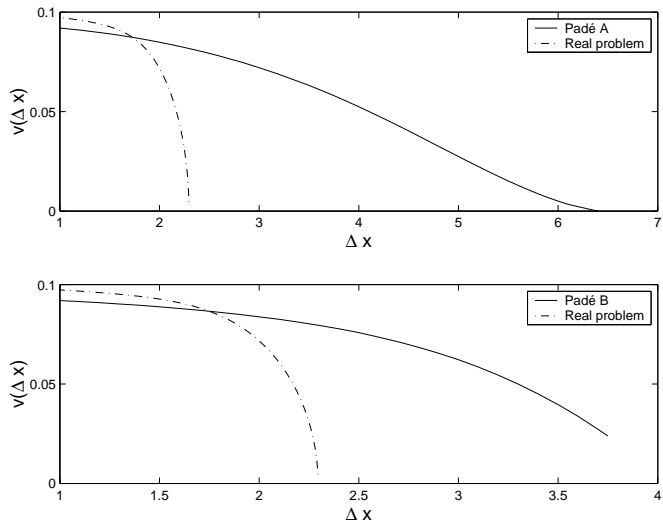  , height=7cm},}
\caption{Effective front speed as a function of $\Delta x$, for the
Pad{\'e} model A (top panel) and model B (bottom panel).}
\lbfig{pfig1}
\end{figure}
\begin{figure}[htbp]
\centering
{\psfig{file=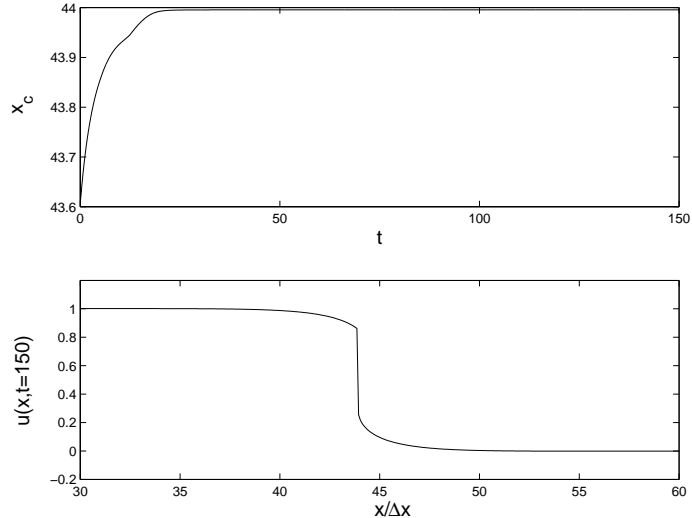  , height=7cm}}
\caption{The figure shows the time evolution of the front for
model A and for $\Delta x=6.4$. The top panel shows the time 
evolution of the front center which eventually leads to trapping.
The bottom panel shows the final front configuration of the numerical
simulation at $t=150$.}
\lbfig{pfig2}
\end{figure}

Our results of this quasi-continuum approach to the discrete problem
can be summarized in \fig{pfig1} and \fig{pfig2}.
\fig{pfig1} shows the speed of the fronts in 
Pad{\'e} models A and B respectively. We can observe that the
critical value of $\Delta x$ beyond which trapping of the front
occurs is significantly displaced from the actual one of $\Delta x^*
\approx 2.3$, for $\eta=0.45$. In particular, for model A, 
$\Delta x^* \approx 6.4$, while for model B, the corresponding
critical value is $\Delta x^* \approx 3.8$. We can deduce that
the latter model is closer to the actual physical reality, even 
though the relevant prediction is still considerably higher
than its actual value for the discrete model. 

In part at least, these results (and the discrepancy from the actual discrete
case) can be justified by observing \fig{pfig2}.
The bottom panel of the figure suggests that the {\it only} way in
which the front can stop in these quasi-continuum Pad{\'e} approximations
is by becoming practically a vertical shock-like structure. In this
case, the ``mass'' of the front which is given by $\int_{-\infty}^{\infty} 
u_x^2 dx$ (see e.g., \cite{boesch} and references therein) becomes
practically infinite. 
This means that the inertia of the front becomes too
big for the front to move and hence ``pinning'' occurs. However,
notice that this process of pinning is significantly different
than the details of the discrete structure of the problem (such
as e.g., the saddle-node bifurcation and the transition to pinned
solutions). The translationally invariant quasi-continuum Pad{\'e}
approximations of models A and B do not ``see'' such features. Instead,
they incorporate the well-known feature of front steepening 
for stronger discreteness \cite{PK} and the criticality of the latter
feature eventually leads to pinning. 

An additional pointer to the
fact that such (pseudo-differential operator) models are ``eligible''
to pinning is that they are devoid of some of the important symmetries
that are inherently related to traveling such as the Galilean 
invariance in the case of continuum bistable equation or the 
Lorentz invariance of its Hamiltonian (nonlinear Klein-Gordon) analog.

\section{Summary and Discussion}

We presented a computer-assisted approach for the {\it solution}
of effective, translationally invariant equations for spatially
discrete problems without deriving these equations in closed form.
Assuming that such an equation exists, its time-one map is approximated
through the coarse time stepper, constructed through an ensemble of
appropriately initialized simulations of the detailed discrete problem.
Combining the coarse time stepper with matrix-free based 
numerical analysis techniques, e.g. contraction mappings 
such as RPM, can then help analyze the unavailable effective equation.
We are currently exploring the use of our coarse time stepper with
coarse projective integration \cite{gk,GKT,Ramiro}.
Matrix-free eigenanalysis techniques should also be explored, 
especially since they can help test the ``fast slaving" hypothesis 
underlying the existence of a closed effective equation (see, for
example, the discussion in \cite{Makeev1,Hummer}).

We also presented initial computational results exploring the
effect of certain ``construction parameters" of the approach: the
number of shifted copies in the ensemble of initial conditions, as well as
the time-horizon used.
We included a comparison between our approach and a particular way
of obtaining explicit approximate translationally invariant 
evolution equations for such a problem (the Pad{\'e} approximation).
More work is necessary along these lines, exploring the relation of our approach
with traditional homogenization methods at small lattice spacings.
A discrete problem whose detailed solution can be obtained
explicitly (perhaps a piecewise linear kinetics problem) or at least
approximated very well analytically over short times, would be the
ideal context in which to study these issues.

Several extensions of the approach can be envisioned, and might
be interesting to explore.
A time stepper based approach can be applied without modification
to hybrid discrete-continuum media, e.g. continuum transport with
a lattice of sources or sinks, such as cells secreting ligands into
and binding them back from a liquid solution, \cite{Stas}.
It is clear that it can be tried in more than one dimensions, and for
regular lattices of different geometry.
For irregular lattices the
averaging ``over all shifts" we performed here for periodic media
can be substituted with a Monte Carlo sampling over the distribution
of possible lattices that takes into account what we know about the
statistical geometry of the lattices. 
%
In this paper we assumed that an equation existed and closed for 
the {\it expected shape} of the solution.
Conceivably one can 
attempt to develop time steppers not only for the expectation
(the first moment of a distribution of possible results), but, say,
for the expectation {\it and} the standard deviation of possible
results; the lifting operator would then have to be appropriately 
modified.
Finally, our time stepper here was built on short simulations of 
the {\it entire} detailed discete system in space. 
Hybrid simulations, where a known, explicit effective equation 
is accurate over {\it part} of the physical domain can be done;
an ``overall hybrid coarse'' 
time stepper (explicit equation over part of the domain,
and the coarse time stepper in this paper 
over the rest of the domain) will then be used.
In a multiscale context, we have proposed ``gaptooth" and ``patch
dynamics" simulations \cite{GearPatch,Manifesto}, 
where the present coarse time stepper integrations are
performed not over the entire domain, but over a mesh of small 
computational ``boxes". 
Both hybrid and ``gaptooth" simulations, if possible, require careful
boundary conditions for the ``handshaking" between the continuum
equation and the discrete simulations, or the discrete simulations
in distant boxes, effectively implementing smoothness of the solution
of the unavailable effective equation (e.g 
\cite{Manifesto,LiJu1,LiJu2,Philips,Weinan}).

We close with a discussion of the ``onset of pinning", the transition
around which our test example of the coarse time stepper was focused.
Continuum effective equations such as the ones discussed here through the
numerical time-stepping procedure do {\it not}, strictly 
speaking, possess a bifurcation at the critical point of the 
genuinely discrete problem. 
In this effective process, the
bifurcation is smeared out and rendered a ``continuum transition''
(see, for example, materials science models of the onset of
movement of a front, \cite{Cahn,Maroudas}).
On the other hand, one might argue that this is an acceptable,
and possibly optimal way for a
continuum equation to represent the discrete bifurcation to pinning. 
We can see that other procedures, such as the discreteness-emulating
Pad{\'e} type ones, lose a lot of the quantitative structure of the
relevant transition. 
On the other hand, if a continuum {\it
differential} (as opposed to pseudo-differential) equation
was constructed to ``model'' this transition, the latter would
possess other artificial features such as a topologically mandated,
unstable branch of traveling wave solutions \cite{nicolaenko}.
It is conceivable that the short hysteresis loop sometimes predicted
by the coarse time stepper close to pinning conditions is a ``vestige"
of this unstable branch that translationally invariant equations would
necessarily predict.
In conclusion, it can be appreciated that genuinely discrete problems
and continuum ones have inherent differences\footnote{A similar example
can be found in the comparison of discrete and {\em periodic} 
continuum problems,
where the former ones possess a single
permissible band of excitations, while the latter possess an 
infinity of such bands and hence allow for interband transitions 
\cite{wannier}.} that cannot be fully captured by emulating (or 
``summarizing'') the one context through the other. Nevertheless, the approach
proposed here, combined with a ``common sense'' interpretation
of its results with respect to the genuinely discrete problem,
performs in a satisfactory way for the modeler,
even for the ``most different'' features between
discrete and continuum models.

\section*{Acknowledgments}

Part of the research for this paper was carried out while
Olof Runborg held a post-doctoral appointment with the Program
for Applied and Computational Mathematics at Princeton University, supported
by NSF KDI grant DMS-9872890.
Panayotis G. Kevrekidis gratefully acknowledges
support from a UMass FRG, NSF-DMS-0204585 and from the Eppley Foundation
for Research.
Kurt Lust is a postdoctoral fellow of the Fund for Scientific Research---Flanders. This paper presents research results of the Belgian Programme on Interuniversity Poles of Attraction, initiated by the Belgian State, Prime Minister's Office for Science, Technology and Culture. The scientific responsibility rests with its authors. 
Ioannis G. Kevrekidis 
gratefully acknowledges the support of AFOSR (Dynamics and Control)
and an NSF-ITR grant.

\end {document}